\magnification1200 \rightline{KCL-MTH-06-08, UB-ECM-PF 06/22,
TIT/HEP-555, Toho-CP-0682} \rightline{hep-th/0607104}

\vskip .5cm \centerline {\bf Diffeomorphism, kappa transformations
and the theory of non-linear realisations}

\vskip 1cm \centerline{Joaquim Gomis}
\centerline{Departament ECM, Facultat F\'{\i}sica}
\centerline{Universitat Barcelona, Diagonal 647} \centerline{E-08028
Barcelona, Spain} \centerline{Department of Physics}
\centerline{Tokyo Institute of Technology} \centerline{Meguro, Tokyo
152-8551, Japan} \centerline{} \centerline{Kiyoshi Kamimura}
\centerline{Department of Physics}\centerline{Toho University,
Funabashi 274-8510, Japan} \centerline{} \centerline {and}
\centerline{} \centerline{Peter West} \centerline{Department of
Mathematics} \centerline{King's College, London WC2R 2LS, UK}

\vskip 0.5cm \vskip .2cm \noindent We will show how the theory of
non-linear realisations can be used to  naturally incorporate world
line diffeomorphisms and kappa transformations for the point
particle and superpoint particle respectively. Similar results also
hold for a general p-brane and super p-brane, however, we must in
these cases  include an additional Lorentz  transformation. \vskip
.5cm

\vfill \eject

{\bf 1 Introduction}
\medskip
 Non-linear realizations played an important role in particle
physics. 
It has first been  developed in constructing lagrangians with
chiral symmetries [1]. 
In this paper we will consider how to use
the theory of non-linear realisations to derive the dynamics  and
symmetries of  branes. As such, it will be instructive to summarise
the theory of non-linear realisations in general. For the case of
internal symmetries, the original general formulation [2] of the
non-linear realisation considered a group $G$  with sub-algebra $H$
and worked with elements of the coset $G/H$ which depended on
space-time $x$.  Specifically, they considered coset representatives
in the form of a set of  elements $g(x)$ of $G$. Under a rigid
transformation of the group $g_0$ the coset representatives $g(x)$
changed as $g(x) \to g_0 g(x) $. However, in order to preserve the
choice of coset representatives one had to simultaneously performed
a specific transformation that belonged to $H$ and depended on $g_0$
and $g(x)$. Hence the coset representatives changed as
$$
g(x)\to g_0 g(x) h(g_0, g(x))\;. \eqno(1.1)$$
The non-linearly realised
theory is taken to be the one invariant under this symmetry
transformation. To be specific  it is often useful   to take the
group elements to be in exponential parameterisation and choose the
coset representatives to be $exp (i\phi_a (x)  T^a)$ where $T_a$ are
the generators of $G$ which are not in $H$ and $\phi_a(x)$ will
become the fields of the non-linearly realised theory.
\par
Non-linear realisations were also used  [3]  to derive symmetries
that involved transformations of manifolds that were to be
interpreted as space-times. That is the manifolds are the coset
space $G/H$. The choice of coset representatives then provides a
parameterisation of the manifold. The classic case is that of
superspace [4] where the group $G$ is the supersymmetry group
involving the translation and supercharge generators, denoted
generically by $P, Q$ and Lorentz transformations and the subgroup
$H$ is just the Lorentz  group. The coset representatives may be
chosen to be  of the  form $g(x,\theta) =exp(i xP+\bar\theta Q)$ where
$(x,\theta)$ are interpreted as   the coordinates  of superspace.
The transformations on superspace being  given by
$$g(x,\theta)\to g_0 g(x,\theta) h(g_0, x, \theta)
\eqno(1.2)$$
with $ h(g_0, x, \theta)$ being the corresponding
compensating Lorentz transformation. The formulae for any such
non-linear realisation are  analogous.
\par
We can also have non-linear realisations that concern symmetries
that act on space-time and on the fields that live on it [3]. In
this case we consider a group $K$ with a sub-algebra $H$ and a set
of generators that belong to a representation $L$ of $K$. Together
$K$ and $L$ form a semi-direct product group denoted $G=K\otimes_s
L$, that is the commutator of a generator of $K$ and one of $L$ is a
generator of $L$.  The group elements $l\in L$ will introduce
space-time into theory and if they are parameterised by $x$ these
become the  coordinates of the space-time which is the coset
$H\otimes_s L/H$. We consider the non-linear realisation based on
the coset $G/H$. The corresponding group element used in the
non-linear realisation takes the form $k=l(x) g(x) $ where $l(x)\in
L$,  and $g(x)$ is a coset representative of $K/H$. The coset
representatives $g(x)$ are taken to depend on these coordinates $x$
and they will contain the fields that live on the space-time. Under
a group transformation of $K\otimes_s L$,
 $g(x)\to k_0 g(x) h(k_0,l,g(x))$ where $k_0\in G$
 and the latter factor  is the correcting $H$
transformation required to preserve the  choice of coset
representatives in $K/H$. If $k_0\in L$ then the transformation will
just change the coordinates $x$, however, if $k_0 \in K$ then the
transformation will change the coordinates $x$ and also the fields
contained in $g(x)$.  One of the first examples of this type of
non-linear realisation was for the conformal group in reference [5].
\par
Although the original papers [2] of Callan, Coleman, Wess and
Zumino, and much of the subsequent development, were formulated in
terms of coset representatives it is more elegant to reformulate the
theory of non-linear realisations in terms of the group elements of
$G$ themselves rather than just the coset $G/H$.  We first discuss
this approach [6,7] for the case of an internal symmetry where we
work with the group elements of $G$ and take the symmetry of the
non-linearly realised  theory to be by definition given by
$$
g(x)\to g_0 g(x) ,\ \ {\rm and }\ \  g(x)\to g(x) h(x)
\eqno(1.3)$$
where $g_0\in G$ is a rigid transformation while  the second
independent  local transformation $h(x)$ is an
arbitrary space-time dependent
transformation that belongs to $H$. In this more general way of
doing things the theory contains fields that  can be fixed
using the local $H$ transformations. We will
refer to such a formulation as a non-linear realisation of an group
$G$ with local subgroup $H$.  One can from the beginning  use these
local $H$ transformations to set to zero some fields and so work
only with a coset representatives. This is equivalent to the
original approach and it requires the compensating transformations
discussed above. One can also work in a half way house where only
some of the fields are removed. This  more general
approach to formulating  non-linear realisation can be used for
space-time symmetries as well as internal symmetries of
fields.  In all future discussions in this
paper we will adopt this more general way of constructing non-linear
realisations although we will often use some of the local symmetry
to fix certain less interesting parts of the group element.
\par
One can view branes   solitonic objects that occur in field theories
and from this perspective their motion can be seen as a non-linear
realisation corresponding to the symmetries that are spontaneously
broken by the soliton. The embedding coordinates transverse to the
brane are the Goldstone bosons for the broken translations and in a
supersymmetric theory part of the fermions correspond to the
breaking of the supersymmetries by the solitons. Some early papers
on the derivation of the  dynamics of specific supersymmetric branes
from the viewpoint of non-linear realisations are given in [8-12].
In general we must begin with a group $K$, with a specified subgroup
$H$, together with  a representation $L$ which form a semi-direct
product group denoted $G=K\otimes_s L$. We then further sub-divide
the generators in the representation $L$ of $K$ into two sets $L_1$
and $L_2$ which are representations of $H$, the specified sub-algebra
of $K$.   We may think of the algebra $K$ as an automorphism algebra
of the algebra $L$, which is not always a commuting algebra.  The
local subgroup of the non-linear realisation has in the past been
taken to be the group $H$. This procedure is most easily explained
by considering the case which leads to the dynamics of the bosonic
p-brane [13]. To do this we take $K=SO(1,D-1)$ and $L$ are the
translation generators in the vector representation of
$K=SO(1,D-1)$.
 We then take $L_1$ to contain $p +1$ of the translation generators
and $H=SO(1,p)\otimes SO(D-p-1)$. The generators of $L_1$ being a
vector under the first factor and a scalar under the second factor
of $H$. The sets  $L_1$ and $L_2$ then contain the unbroken and
broken translations and  $H$ the unbroken Lorentz rotations. The
coset representatives can be generically written as
$$
g=e^{i x\cdot P}e^{i x^\prime\cdot P^\prime }e^{i \phi\cdot
J^\prime} \eqno(1.4)$$ where $P$, $P^\prime$ belong to $L_1$ and
$L_2$ respectively and $J^\prime$ are the generators of $K$ not in
$H$.
\par
Even in this apparently simple example  there are still several
different possible options that can be pursued. The first  option
concerns the Goldstone fields corresponding to broken Lorentz
transformations. In reference [13] these were algebraically
eliminated in terms of derivatives of the Goldstone fields for
translations, using a constraint that was covariant under the
non-linear realisation, that is a so called inverse Higgs constraint
[14].  After this elimination  an action was constructed. However,
recently a new action for the bosonic p-brane was found that was  a
function of all the above Goldstone fields including those
corresponding to broken Lorentz transformations [15]. The equations
of motion for these latter Goldstone bosons were in fact non-other
than the inverse Higgs constraint and using this in the action one
found the previous  result.
\par
The second option is of a more fundamental nature and it concerns
the way the non-linear realisation is defined from the outset. In
particular, it   concerns the dependence of the fields. It was usual
in discussions of brane dynamics to take  the unbroken translations
that occurred in the group element to be associated with
coordinates, which turned out to parameterise the brane world, and
the  Goldstone fields were taken to depend on these coordinates.
However,  reference [13] also introduced  fields associated with the
unbroken translations which appeared in the group element and these
in common with all the Goldstone fields were taken to depend on
external parameters $\xi$  that turned out to parameterise the brane
world volume. It was then demanded that the theory be
reparameterisation invariant as a requirement  in addition to be
invariant under the transformations of the non-linear realisation of
equation (1.3).  By choosing static gauge in the latter formulation
one can recover the former formulation. Non-relativistic branes for
Galilei [16] and Newton Hooke [17] groups have been constructed
using non-linear realizations along the lines of the  previous
paragraph. The corresponding actions are Wess-Zumino terms of the
previous groups.
\par
In this paper we will consider how to incorporate the local
symmetries such as world volume reparameterisation and, for
supersymmetric branes, $\kappa$-symmetry into the theory of
non-linear realisations. Although reference [13] had the advantage
that it introduced the reparameterisation invariance in the
parameters $\xi$ to be present in brane dynamics,  it also
introduced fields corresponding to generators that were not broken.
The resolution of this puzzle is to adopt the formulation of the
theory of non-linear realisations encoded in the transformations of
equation (1.3), that is work with the group elements rather than the
just the coset,  and take the unbroken translation generators to be
part of the local subgroup of the non-linear realisation. Thus the
unbroken translation generators occur in the group element with
fields, which depend, like all other fields, on the parameters
$\xi$, However, as the unbroken translations are in the local
subgroup their corresponding fields can be fixed by a suitable local
transformation. The transformations of the non-linear realisation
corresponding to the unbroken translations should then correspond to
the reparameterisation invariance of  the theory. For the point
particle we will find that this is indeed the case. However, for the
general bosonic p-brane  we will find that we must include with
these local transformations an additional Lorentz rotation.  In this
paper will adopt the latter point of view of reference [15] in that
we will find actions before the elimination of the Goldstone fields
corresponding to the broken Lorentz transformations.
\par
In section three we will consider the super p-brane. In this case we
not only have diffeomorphism invariance, but also kappa
transformations. The discussion follows a similar path to that given
in section two  for the    bosonic case. Now the local subgroup will
include  the unbroken translations and also the unbroken
supersymmetries. The corresponding  local transformations will lead
for the superpoint particle to world line diffeomorphisms and kappa
supersymmetry transformations respectively. However,  for the
general super p-brane the same results will hold  except that one
must include an additional Lorentz rotation. The derivation of these
results depend crucially on having Goldstone bosons associated with
the broken Lorentz transformations. For the case of super p-branes
we will also use the ideas of [15] and not introduce any
superfields. In particular, we will take the fields to just depend
on the parameters $\xi$ and not on any additional Grassman odd
parameters. As explained in reference [15], in a non-linear
realisation all the fields are associated with generators of the
algebras involved  and so appear automatically in the group element
with their corresponding generator.

\par

\medskip
{\bf 2 The Point Particle and Bosonic  Brane}
\medskip
We wish to  construct the non-linear realisation that leads to
dynamics of  the bosonic p-brane moving in a $D$-dimensional
space-time and so consider the algebra of translations and Lorentz
rotations $ISO(1,D-1)$
$$
[J_{\underline a\underline b}, J_{\underline c\underline d}]=
-i\eta_{\underline b\underline c}J_{\underline a\underline d}
+i\eta_{\underline a\underline c}J_{\underline b\underline d}
+i\eta_{\underline b\underline d}J_{\underline a\underline c}
-i\eta_{\underline a\underline d}J_{\underline b\underline c}
\eqno(2.1)$$
$$
[J_{\underline a\underline b}, P_{\underline c}]=-i \eta_{\underline
b\underline c}P_{\underline a}+ i\eta_{\underline a\underline
c}P_{\underline b}\;. \eqno(2.2)$$
In terms of our above discussion in
the introduction $K=SO(1,D-1)$ and $L$ contains the translations
$P_{\underline a}, \ \underline a=0,1,\ldots , D-1$. We sub-divide
the latter into $L_1=P_{ a}, \ a=0,\ldots , p$ and $L_1=P_{ a'}, \
a'=p+1,\ldots , D-1$ and take the local sub-algebra of the
non-linear realisation to be $H=ISO(1,p)\otimes SO(D-p-1)$ which
contains the generators $ P_a, J_{ab}, J_{a'b'}$. The $P_a$
correspond to the unbroken translations and $ J_{ab}, J_{a'b'}$ the
unbroken Lorentz rotations. We are using the notation that an
underlined index, i.e. $ \underline a$ goes over all possible values
from $0$ to $D-1$ while the unprimed indices $a$ take the values
$a=0,\ldots , p$ and primed indices $a'$ take the values $p+1,\ldots
, D-1$. The latter are the indices which are longitudinal and
transverse to the brane respectively. The new feature is that we
have placed the unbroken translations in the local sub-algebra of
the non-linear realisation.
\par
We now construct  the  non-linear realisation with algebra
$G=ISO(1,D-1)$ with local sub-algebra $H=ISO(1,p)\otimes SO(D-p-1)$.
We take the group element to be given by
$$
g=e^{ix^a P_a} e^{ix^{a'} P_{a'}}e^{i\phi_{a}{}^{b'}  J^{a}{}_{b'}}
\;.\eqno(2.3)$$ We note that this is not the most general line element
as we have used the local Lorentz transformations to set part of the
group element to one. All the above fields are taken to depend on
the parameters $ \xi^i, \ i=0,\ldots , p$. As such the procedure
has some aspects in common with the case of an internal symmetry.
The non-linear realisation is by definition a theory that is
invariant   under the transformations of equation (1.3) which in
this case reads
$$
g(\xi)\to g_0 g(\xi) ,\ \ {\rm and }\ \  g(\xi)\to g(\xi) h(\xi)
\;.\eqno(2.4)$$
\par
We are particularly interested in local unbroken translations which
are the new feature compared to other treatments of the bosonic
brane and can be used to fix the field $x^a$ that appears in the
group element together with the unbroken translations $P_a$. As such
we can consider local transformations of the form $h=e^{ir^a P_a}$
where $r^a $ is an arbitrary function of $ \xi$. Carrying out the
transformation $ g(\xi)\to g(\xi) h(\xi)$ we find that the fields
transform as
$$
\delta x^{n}=r^b (\Phi^{-1})_b{}^{n}\equiv s^{n},\ \delta x^{n'}=r^b
(\Phi^{-1})_b{}^{n'},\ \delta \phi_{a}{}^{b'}=0 \eqno(2.5)$$ where
for an arbitrary  Lorentz transformation involving only the broken
Lorentz generators $J^{a}{}_{b'}$   we define
$$
e^{-i\phi\cdot J}P_{\underline a}e^{i\phi\cdot J}=\Phi_{\underline
a}{}^{\underline b} P_{\underline b}
\;.\eqno(2.6)$$
It is convenient
to take use an explicit form of the Lorentz transformation
$\Phi_{\underline a}{}^{\underline b} $.  We may  write it in the form
$$
\Phi= \left(\matrix{I_1& \varphi \cr -\varphi^T &
I_2\cr}\right)\left(\matrix{B_1& 0 \cr 0 &
B_2\cr}\right)=\left(\matrix{B_1& \varphi B_2 \cr -\varphi^TB_1 &
B_2\cr}\right)
\eqno(2.7)$$
where
$B_1=(I_1+\varphi\varphi^T)^{-{1\over 2}}$ and
$B_2=(I_2+\varphi^T\varphi)^{-{1\over 2}}$. The inverse Lorentz
transformation is given by
$$
\Phi^{-1}=\left(\matrix{B_1& -B_1\varphi  \cr B_2\varphi^T &
B_2\cr}\right) \;.\eqno(2.8)$$
The broken Lorentz Goldstone fields ${\phi_a}^{b'}$ in (2.3) and
$\varphi_{a}{}^{b'}$ are related by
$${(\varphi)_{a}}^{b'}= {\phi_a}^{c'}{({{\sinh
\tilde V}\over{\tilde V\cosh \tilde V}})_{c'}}^ {b'},\qquad
{(\tilde V^2)_{a'}}^{b'}\equiv -\phi^{c}{}_{a'}\phi_{c}{}^{b'}.
\eqno(2.9)$$
For more details on this particular
parameterisation of the Lorentz transformation
we refer the reader
to reference [15]. In terms of this parameterisation the variations of
equation (2.5) can be written as
$$
\delta x^{n}=r^b (B_1)_b{}^{n}\equiv s^{n},\ \delta x^{n'}=-r^c
(B_1)_c{}^b\varphi_b{}^{n'}=-s^b \varphi_b{}^{n'}
\;.\eqno(2.10)$$
\par
The dynamics is  built out of the Cartan forms ${\cal
V}=-ig^{-1}dg= {\cal V}_id\xi^i$
$$
{\cal V}_i= g^{-1}\partial_ig= e_i{}^a P_{a}+e_i{}^{a'} P_{a'}+{1
\over 2} w_{i \underline a}{}^{\underline b}   J^{\underline
a}{}_{\underline b} \eqno(2.11)$$
which are inert under rigid $G$
transformations and transform under local $ H $ transformations as
$$
{\cal V}_i\to h^{-1}{\cal V}_i h-i h^{-1}{\partial}_i h \;.\eqno(2.12)$$
The Cartan forms are given by
$$
e_i{}^a= (\partial_i
x^c-\partial_ix^{b'}\varphi^T_{b'}{}^{c})(B_1)_c{}^a,\
e_i{}^{a'}=(\partial_i x^c \varphi_{c}{}^{b'}+\partial_i
x^{b'})(B_2)_{b'}{}^{a'},\  w_{i\underline a}{}^{\underline b}
=(\Phi^{-1})_{\underline a}{}^{\underline c}\partial_i (\Phi)_
{\underline c}{}^{\underline b} \;.\eqno(2.13)$$ $e_i{}^{\underline {n}}$
is the vielbein and $w_{i\underline a}{}^ {\underline b}$ is the
spin connection. The transformations of the Cartan forms under the
local transformations corresponding to $P_a$ are given by
$$
\delta e_i{}^a=\partial_i r^a-r^bw_{ib}{}^{a},\quad \delta
e_i{}^{a'}=-r^bw_{ib}{}^{a'},\quad \delta w_{i\underline
a}{}^{\underline b}=0 \;.\eqno(2.14)$$
\medskip

{\bf Bosonic point particle }
\medskip
We now  consider  the case of the point particle, that is $p=0$.   Although
this is a very elementary system it will allow us to illustrate how the
world-line reparameterisation arises from the non-linear realisation by
taking the unbroken time translation to be part of the local subgroup.
To have a more usual notation we take $\tau=\xi^{0}$ and $t=x^{0}$. The
Cartan forms of equation (2.13) are given by
$$
e\equiv e_0{}^{0}= ({dt\over d\tau}-{d x^{b'}\over d\tau}\varphi_{b'})
(1+ \varphi^{b'}\varphi_{b'})^{-{1\over 2}},\   e^{a'}\equiv
e_0{}^{a'}=({d  t\over d\tau} \varphi^{b'}+{d x^{b'}\over
d\tau})(B_2)_{b'}{}^{a'},
$$
$$ w_{\underline
a}{}^{\underline b}\equiv  w_{0\underline a}{}^{\underline b}
=(\Phi^{-1})_{\underline a}{}^{\underline c}{d\over
d\tau}(\Phi)_{\underline c}{}^{\underline b} \eqno(2.15)$$ where
$\varphi_{0}{}^{b'}\equiv \varphi^{b'}, \varphi^{0}{}_{b'}\equiv
\varphi_{b'}=-\varphi^{b'}$. Under a local $P_{0}$ transformation the
fields transform as
$$
\delta t=r (1+ \varphi^{b'}\varphi_{b'})^{-{1\over 2}}\equiv s,\quad
\delta x^{n'}=-s \varphi^{n'} , \eqno(2.16)$$ where $r=r^{0}$ and
$s=s^{0}$, while the  Cartan forms transform associated with the
translations as
$$
\delta e={ dr\over d\tau},\quad \delta e^{a'}=- r{w_{0}}^{a'}
,\quad \delta w_{\underline a}{}^{\underline b}=0 . \eqno(2.17)$$

\par
Examining the first of equations in (2.17) we realise that an action
invariant under  local $P_{0}$ transformations, and indeed all the
transformations of equation (2.4), is given by
$$
A=-\int d\tau e=-\int d\tau {({dt\over d\tau}-{d x^{b'}\over
d\tau}\varphi_{b'})\over (1+ \varphi^{b'}\varphi_{b'})^{{1\over 2}}}
\;.\eqno(2.18)$$ Varying this action we find the equation of motion for
$\varphi^{a'}$ implies that
$$
\varphi^{a'}=-{dx^{a'}\over d\tau}\slash{dt \over d\tau}
\;.\eqno(2.19)$$ Substituting this algebraic equation back into the
action we find the standard action for the point particle, namely
$$
A= -\int d\tau (-{dx^{\underline {n}}\over d\tau} {dx_{\underline
{n}}\over d\tau} )^{{1\over 2}} \;.\eqno(2.20)$$
Taking the  equation of motion of $\varphi^{a'}$, that is
implementing equation (2.19), corresponds to setting the Cartan form
$ e^{a'}$ of equation (2.15) to zero, i.e.
$$ e^{a'}=0
\;.\eqno(2.21) $$ We see, using equation (2.17), that   this condition
is not invariant under the transformations of the non-linear
realisation, but instead implies the equation of motion of $x^{\underline a}$,
 $w_{0}{}^{a'}=0$,  for the
point particle. That it  means an   equation of motion
follows from the fact that $e^{a'}=0$ is a consequence of the above
invariant action and so all the equations of motion of the action
must rotate into each other. We note that if  one does not take the
local sub-algebra to contain the unbroken translations, then  setting
the Cartan form $ e^{a'}=0$ is an invariant condition and so is an
example of the inverse Higgs effect. This latter  point of view and
the above action are the subject of reference [14].
\par
Finally, we examine the transformations of the fields if equation
(2.19) holds. The transformation of $t$ of equation (2.16) can be
considered as a transformation of $\tau$ by identifying $\delta t=s=
{d t\over d\tau}\rho$ where $\delta \tau =\rho$. Using equations
(2.16), (2.17) and (2.19) we find that
$$
\delta x^{n'}= {dx^{n'}\over d\tau} \rho,\quad \delta e={d\over d\tau}
(e \rho) \eqno(2.22)$$ which we recognise as the standard variations
under a diffeomorphism in $\tau$. In fact even before one implements
$ e^{a'}=0$,   one finds that  the transformations in (2.16)
differ from  those of equation (2.22) by anti-symmetric combinations of
equations of motion. We have therefore found that implementing the
$\varphi^{a'}$ equation of motion converts the local $P_{0}$   transformation
of the non-linear realisation into a diffeomorphism of the world-line.
\medskip
{\bf The  bosonic p-brane}
\medskip
We now consider the bosonic p-brane which we will see   requires an
additional step compared to that for the point particle. An
action that has been derived from the theory of non-linear   realisations in
the absence of local unbroken translation and depends on all
the Goldstone bosons, including those for broken Lorentz
transformations is given by [15]
$$
{\cal L}=-{1\over{(p+1)!}}\epsilon_{a_0,...,a_p}e^{a_0}\wedge
.... \wedge e^{a_p}=
d^{p+1}\xi \det e_i{}^a
\;.\eqno(2.23)$$
Here $ e{}^a=d\xi^ie_i{}^a$ is the Cartan form associated to $P_a$ of
equation
(2.13).
\par

This action is a functional of $\varphi_a{}^{n'}$ and
$x^{\underline n} $ and varying with respect to  $\varphi_a{}^{n'}$
leads to an equation of motion that implies
$$\varphi_a{}^{n'}=- {{\partial X^{n'}}\over {\partial \xi^i}}
\left({{\partial X^{a}}\over {\partial \xi^i}}\right)^{-1}
=- {{\partial X^{n'}}\over {\partial \xi^i}}
(e^{-1})_a{}^i
\eqno(2.24)
$$
which  is the solution of $e_i{}^{a'}=0$ [15]. Eliminating
$\varphi_a{}^{n'}$ from the action we find the usual action for the
p-brane. Actions for branes including supplementary variables,
Lorentz vector harmonics, were considered in [18]. These actions
were not derived from the view point of non-linear realizations and
contain redundant fields that satisfy constraint equations.

\par
As the action of equation (2.23) is  constructed from the Cartan  forms of
equation (2.11) and  it is invariant under local
$SO(1,p)\otimes SO(D-p-1)$ transformations. Consequently, it is also
invariant under  rigid $SO(1,D-1)$ transformations.
However, under the local
transformations corresponding to local $P_a$ transformations of equation
(2.14) this action transforms as
$$
\delta \int d^{p+1}\xi \det e_i{}^a= \int d^{p+1}\xi (\det e_i{}^a)
\;r^k(e_k{}^{b'}w_{ib'}{}^{a} (e^{-1})_a{}^i- (e^{-1})_a{}^i
e_i{}^{b'}w_{kb'}{}^{a} ) \eqno(2.25)$$ where $r^i=r^a(e^{-1})_a{}^i
$. Although  this does not vanish in general  it does vanish when
$e_i{}^{a'}=0$  and since this is the equation of motion of
$\varphi_a{}^{n'}$ it follows that one can cancel this term by
adding a suitable quantity to the variation of  $\varphi_a{}^{n'}$.
Under a combined local translations with parameter $r^a$ and the
above local Lorentz transformation with parameter $r_a{}^{b'}$  we
find that one form $e{}^{\underline a}$ varies as
$$\delta e^a=dr^a+w^a{}_br^b+e^{b'}r_{b'}{}^a,\qquad
  \delta e^{a'}=e^{b}r_{b}{}^{a'}+w^{a'}{}_br^b.
\eqno(2.26)$$
One can verify that the Lorentz transformation which indeed cancel the
above variation of the action (2.25) is given by
$$
r^{a'}{}_b= -(e^{-1})_c{}^i r^c w_{ib}{}^{a'}+ r_b((e^{-1})_c{}^i
w_{i}{}^{ca'}).
\eqno(2.27)$$
As we  have an invariance of the action the equation of motion
$e_i{}^{a'}=0$ is preserved provided we use   the other equations of
motion,  as must be the case.
\par
It is instructive to consider the transformation of $e_i{}^a$ under a
local translation. Using $r^i=(e^{-1})_a{}^i r^a$ we can express
equation (2.14) as
$$ \eqalign { \delta e_i{}^a&=\partial_i(r^j) e_j{}^a
+r^j (\partial_i e_j{}^a-w_{ib}{}^{a} e_i{}^b) \cr &=\partial_i(r^j)
e_j{}^a+r^j \partial_j e_i{}^a +(r^j{{w_j}^a}_b){e_i}^b +r^j
(w_{ib'}{}^{a} e_j{}^{b'}-w_{jb'}{}^{a} e_i{}^{b'}) \cr}
\eqno(2.28)$$
where in the last step we have used the Maurer-Cartan equations. Thus we
find the standard diffeomorphism plus a local Lorentz rotation plus a
term that vanishes if $e_i{}^{a'}= 0$. Furthermore if we consider a local
translation combined with the Lorentz transformation of (2.26)  we
find that the variation of $e_i{}^a$ is given by
$$ \eqalign { \delta e_i{}^a
&=\partial_i(r^j)
e_j{}^a+r^j \partial_j e_i{}^a +(r^j{{w_j}^a}_b){e_i}^b +r^j
(w_{ib'}{}^{a} e_j{}^{b'}-w_{jb'}{}^{a} e_i{}^{b'})+e_i{}^{b'}r_{b'}{}^a.  \cr}
\eqno(2.29)$$
We note that the first two terms are a diffeomorphism,
the third term is a local $SO(1,p)$ rotation and the remaining terms
vanish when (2.24) is used.

\par
Hence we have found an invariant action which is constructed from
the Cartan forms,  but we have had to modify the  local
transformations from those that follow from the strict application
of the non-linear realisation. The theory is invariant under a left
transformation of the form
$$
h=\exp (ir^a P_a +i r_{a}{}^{b'}J^a{}_{b'}) \eqno(2.30)$$
where $r^{a'}{}_{b}$ is given in equation (2.27).

\medskip
{\bf 3. The Super Point Particle and Super Brane. }
\medskip

In this section we generalize the discussion of the last section on
the bosonic particle and p-brane  to the case of supersymmetric branes.
In particular,  we will study  how the kappa symmetry of
supersymmetric brane  actions arise as local transformations when we
take the unbroken supersymmetry generators as part of the  local
sub-algebra of the non-linear realisation. That kappa
transformations might be part of the non-linear realisation has been
suggested in the past.   For example, in reference [19], the idea of
incorporating local unbroken supersymmetry transformations was
considered and some progress was   made to deriving
$\kappa$-transformations as part of the non-linear realisation.
However,  the author considered the theory to be defined on the
coset and did not introduce Goldstone bosons associated with the
broken Lorentz transformations and so was able to obtain only
partial results.
\par

We take  the algebra $G$ of the non-linear realisation to be the
Super Poincare algebra which contains in addition to equations (2.1)
and (2.2) the commutation relations
$$
\left[Q_{\underline \alpha},~J_{{\underline{a b}}}\right] = -
{{i}\over{2}}  (\Gamma_{\underline{ab}})_{\underline \alpha}{}^
{\underline \beta} Q_{\underline \beta}, \quad \left[Q_{\underline
\alpha},~P_{\underline{a}}\right] = 0, \quad \{ Q_{\underline
\alpha},Q_{\underline \beta}\} = 2(\Gamma^ {\underline
a}C^{-1})_{\underline \alpha\underline \beta} \;P_{\underline {a}}.
\eqno{(3.1)}$$

In order to construct the non-linear realisation we must choose a
local sub-algebra. We take this to have a  Grassman even  part given
by the Lorentz transformations $SO(1,p)\otimes SO(D-p-1)$ and  the
local unbroken translations $P_a$ while the Grassman odd part
contains the supercharges $Q^*$ which are related to the full set of
supersymmetry generators  $Q$ by a projection operator ${\cal P}$ as
$$
Q^*_{\underline \alpha}= {\cal P}_{\underline \alpha}{}^{\underline
\beta}{} Q_{\underline \beta}. \eqno{(3.2)}$$  For the sake of
simplicity we will study the case where half of the supercharges are
unbroken. The non-zero components of $Q^*_{\underline \alpha}$ are
often denoted by $Q_\alpha$ in the literature. The projector must be
such that the chosen algebra must close and in particular
$$
\{ Q^*_{\underline \alpha}, Q^*_{\underline \beta}\}= ({\cal
P}\Gamma ^a C^{-1}{\cal P}^T)_{{\underline \alpha\underline \beta}}
P_a. \eqno{(3.3)}$$

 We write the projection operator ${\cal P}$ as
$$
{\cal P}={1\over2}(1+\Gamma_*),\qquad \Gamma_*{}^2=1, \eqno{(3.4)}$$
(3.3) requires
$$
\Gamma^{a}C^{-1}\Gamma_*{}^TC=\Gamma_*\Gamma^{a},\qquad
\Gamma^{a'}C^{-1}\Gamma_*{}^TC=-\Gamma_*\Gamma^{a'}. \eqno{(3.5)}
$$
For the superparticle, the supercharges are IIA spinors, and
$\Gamma_*$ is , up to sign,  as
$$
\Gamma_*:=\Gamma^{0}\Gamma_{11}. \eqno{(3.6)}
$$
For the super p-branes the supercharges are minimal spinors and they
exist for $p=1,2$ mod $4$ [20]. The $\Gamma_*$ is  , up to sign, as
$$
\Gamma_*:={1\over{(p+1)!}}\epsilon^{a_{0}...a_p}\Gamma_{a_{0}}...
\Gamma_ {a_p}. \eqno{(3.7)}$$
\par
 The formulation being  considered  in this paper applies equally
well to branes which preserve some supersymmetry and those that
break all the supersymmetries. In the latter case there are no
unbroken supercharges and so no corresponding supercharges in the
local subalgebra of the non-linear realisation. This is consistent
with the fact that the branes that break all the supersymmetries
have no kappa invariance.

We may then write the group elements in   the form
$$
g=g_{0}\;U,\qquad g_{0}:=e^{ix^{\underline a} P_{\underline
a}}\;e^{\bar\theta^{\underline \alpha}Q_{\underline \alpha} },\qquad
U:=e^{i\phi_{a}{}^{b'} J^{a}{}_{b'}} \;, \eqno(3.8)$$ where
$\bar\theta^{\underline \alpha}=\theta_{\underline
\beta}C^{\underline \beta\underline\alpha}$ or equivalently
$\bar\theta=\theta^{T}C$. We parametrise the group element $g$, and
so the fields $x^{\underline a} $, $\theta_{\underline \alpha}$ and
$\phi_{a}{}^{b'} $,  to depend on the worldvolume Grassman even
parameters  $\xi^i,i=0,1,\ldots ,p$ [15]. We have used the local
symmetry under the unbroken local Lorentz transformations  to bring
the group element into the above form, however, we have not used the
local transformations involving the unbroken translations or
supersymmetries to fix any of the corresponding fields. The above
differs from the usual  treatment where one usually takes the fields
to depend on the parameters $\xi^i $ as well as a set of Grassman
odd variables $\theta_{\alpha}$. As explained in [15] the
use of Goldstone superfields to describe the dynamics of the
superbranes is not  needed  and leads to a redundancy.

\par
The Cartan   forms are given by
$$
{\cal V}=U^{-1}{\cal V}_{0} U-iU^{-1}d U= e^{\underline
a}P_{\underline a}-i\bar e^{\underline \alpha} Q_{\underline
\alpha}+{1\over2} w^{{\underline a}{\underline b}}J_{{\underline
a}{\underline b }}\;. \eqno{(3.9)}$$ The expression ${\cal V}_{0}$
which appears in the first term is given by
$$
{\cal V}_{0}=-ig_{0}^{-1}dg_{0}=\pi^{\underline a}P_{\underline
a}-i\bar\pi^{\underline \alpha} Q_{\underline \alpha}
\eqno{(3.10)}$$ where $\pi^{\underline a}$, $\bar\pi^{\underline
\alpha}$ are supervielbeins of Poincare superspace which are given
by
$$
\pi^{\underline
a}=dx^{{\underline{a}}}+i\bar\theta\Gamma^{\underline a}
d\theta,\qquad\qquad \bar\pi^{\underline
\alpha}=d\bar\theta^{\underline \alpha} \;. \eqno{(3.11)}$$ Using
equation (2.6) and the  analogous equation for the spinor
representation ${\tilde\Phi_{\beta}}{}^{\alpha}$ of the boost $U$,
$$
U^{-1}P_{\underline a}U= {\Phi_{\underline a}}^{\underline b}
P_{\underline b},\qquad U^{-1}Q_{\alpha}U=
{\tilde\Phi_{\alpha}}{}^{\beta}Q_{\beta}, \eqno(3.12)$$ we find that
$$
e^{\underline a}=\pi^{\underline b}{\Phi_{\underline b}}^{\underline
a}, \qquad \bar e^{\underline \alpha}=\bar \pi^{\underline
\beta}{\tilde\Phi_{\underline \beta}}{}^{\underline \alpha}, \qquad
w_{\underline a}{}^{\underline b}={(\Phi^{-1})_{\underline a}}
^{\underline c}d{\Phi_{\underline c}}^{\underline b},
\eqno{(3.13)}$$ where ${\Phi_{\underline b}}^{\underline a}$ is
given in equation (2.7)  and
$$\tilde \Phi_{\underline
\alpha}{}^{\underline \beta}=\left(\exp ({i\over 2} \phi\cdot
J)\right)_{\underline \alpha}{}^{\underline \beta}= \left(\exp (i
\phi_a{}^{b'} J^a{}_{b'})\right)_{\underline \alpha}{}^{\underline
\beta}= \left(\exp ( {1\over 2}\phi_a{}^{b'}
\Gamma^a{}\Gamma_{b'})\right)_{\underline \alpha}{}^{\underline
\beta}. \eqno{(3.14)}$$ The relationship  between  the spinorial and
the vectorial representations is as usual given by
$$
{\tilde\Phi}\Gamma_{\underline a}{\tilde\Phi^{-1}}=
{(\Phi^{-1})_{\underline a}}^{\underline b} \Gamma_{\underline b}.
\eqno{(3.15)}$$ We note that the expression for $e^{\underline a}$
is the same as that of equation (2.13) except that $dx^{\underline
a}$ is replaced by $\pi^{\underline a}$. The transformations of the
Cartan forms under the local transformations corresponding to $P_a$,
$Q^*$ and $J^a{}_{b'}$ are given by
$$\delta e^a=dr^a+w^a{}_br^b+e^{b'}r_{b'}{}^a+2i(\bar\kappa_*\Gamma^a e),\qquad
  \delta e^{a'}=e^{b}r_{b}{}^{a'}+w^{a'}{}_br^b+2i(\bar\kappa_*\Gamma^{a'}e) ,
\eqno(3.16)$$
$$\delta\bar e^{\underline\alpha}=D\bar\kappa_*^{\underline\alpha}
+{1\over2}\bar e^{\underline\beta}(\Gamma_{ab'}r^{ab'})_
{\underline\beta}{}^{\underline\alpha},\qquad
\delta w^{ab}=w^{ac'}r_{c'}{}^{b} -w^{bc'}r_{c'}{}^{a},\quad
$$
$$
\delta w^{ab'}=dr^{ab'} +w^{ac}r_{c}{}^{b'} -w^{b'
c'}r_{c'}{}^{a},\qquad
\delta w^{a'b'}=w^{a'c}r_{c}{}^{b'} -w^{b'c}r_{c}{}^{a'},\quad
\eqno(3.17)$$where
$\bar\kappa_*=\bar\kappa_*{\cal P}$ is the $Q^*$ transformation
parameter and $D\bar\kappa_*=d\bar\kappa_*-{1\over4}\bar
\kappa_*(\Gamma_{\underline a
\underline b} w^{\underline a  \underline b})\;$
is the Lorentz covariant derivative.
\medskip

As in the bosonic case it will be much easier to treat the case of
the superparticle which is straightforward, while the super p-brane
has an additional step beyond those that arise naturally in the
theory of non-linear realisations. As such we first consider the
case of the superparticle.

\medskip
{\bf Superparticle}
\medskip
For this case the local sub-algebra consists of the generators of
SO(D-1), the  translation $P_{0}$, and half of the supercharges
$Q^*$. The field in the group element of equation (3.8) just depend
on one parameter which we take to be $\tau$. Using equation (2.7) in
equation (3.13), the Cartan forms, for the superparticle are given by
$$
e^{0}= ({\pi^{0}}-{\pi^{b'}}\varphi_{b'}) (1+
\varphi^{b'}\varphi_{b'})^{-{1\over 2}},\quad e^{a'}=({\pi^{0}}
\varphi^{b'}+{\pi^{b'}})(B_2)_{b'}{}^{a'}, \eqno(3.18)$$ and the
remaining Cartan forms ${\bar e}^{\underline\alpha}$ and $
w_{\underline a}{}^{\underline b}$ can also be read off from
(3.13).
\par
Let us first consider the effect of carrying out a local
transformation for the  $P_{0}$ translation.  Taking  $h=e^{ir
P_{0}}$ the fields transform as
$$
\delta x^{\underline a}=r(\Phi^{-1})_{0}{}^{\underline a}, \qquad
\delta\bar\theta^{\underline\alpha}= \delta\phi_{0}{}^{b'}=0,
\eqno(3.19)$$ which is given more explicitly by,
$$
\delta x^{0}=r(1+ \varphi^{b'}\varphi_{b'})^{-{1\over 2}} \equiv
s,\qquad \delta x^{a'}=-s \varphi^{a'}, \eqno(3.20)$$ while the
Cartan forms transform as
$$
\delta e^{0}={ dr},\quad \delta e^{a'}= -r{w_{{0}}}^{a'},\quad 
\delta w_{\underline a}{}^{\underline b}=0,\quad \delta\bar
e^{\underline\alpha}=0 . \eqno(3.21)$$
\par
Next we  consider the local transformations generated by the
unbroken supersymmetries. The local transformation   is given by
taking $h=e^{\bar\kappa_*Q^*}$, where the $\bar\kappa_*^{\underline
\alpha}= (\bar\kappa_*{\cal P})^{\underline \alpha}$ are 16 independent
gauge parameters that depend on $ \tau$. The transformations of the
fields are given by
$$
\delta x^{\underline a}=i\delta\bar\theta\Gamma^{\underline
a}\theta, \qquad \delta\bar\theta^{\underline \alpha}=
\bar\kappa_*^{\underline \beta} \tilde\Phi^{-1}_{\underline
\beta}{}^ {\underline \alpha},\qquad \delta\phi_{0}{}^{b'}={0}.
\eqno{(3.22)}$$ The corresponding transformations of the Cartan
forms are
$$
\delta e^{\underline a}=
2i(\bar\kappa_*\Gamma^{\underline a}e),\qquad \delta \bar
e^{\underline\alpha}=(D\bar\kappa_*)^{\underline\alpha},\qquad
\delta w_{\underline a}{}^{\underline b}={0}. \eqno{(3.23)}$$
\par
The Cartan forms are inert under the rigid transformations and only
transform under the local part of the non-linear realisation and so
to construct an action out of the Cartan forms we need only take
into account of  the effect of the local subgroup. Let us  first
construct an action invariant under local $ P_{0}$ translation and
local Lorentz transformations.  One such term in the action is the
one form $e^{0}$, as in the purely bosonic particle case. The
presence of an additional term is due to the existence of a
non-trivial Chevalley Eilenberg cohomology of the Super Poincare
group [21]. In particular, there exists  a closed two form $-i\bar
e^{\underline\alpha}\bar e^{\underline\beta}
(\Gamma_{11}C^{-1})_{\underline\alpha\underline\beta}$ that is
invariant  under local $P_{0}$  transformations and can be written
as
$$
-i\bar e^{\underline \alpha}\wedge \bar e^{\underline \beta}
(\Gamma_{11}C^{-1})_{\underline \alpha\underline \beta}
=id\bar\theta\Gamma_{11}d\theta=id(\bar\theta\Gamma_{11}d\theta).
\eqno{(3.24)}$$ It follows that $i\bar\theta\Gamma_{11}d\theta$ must
transform as a $\;d\;$ of something under a local $P_{0}$
transformation.
\par
Hence, assuming the action depends only on velocities, we may
take for our action
$$
A=-\int \left({\cal L}_{NG}+b {\cal L}_{WZ}\right) =-\int \left(e^{0}
+i\,b\, \bar\theta\Gamma_{11}{{d\theta}}\right ). \eqno{(3.25)}$$
The $b$ is an arbitrary real  constant that is not fixed by
demanding invariance under the unbroken local Lorentz and $P_{0}$
translations. It is fixed once
the invariance under the local unbroken supersymmetry is imposed.
Using (3.22) and  (3.23) the variation of the lagrangian under these
transformation is given by
$$
\delta_\kappa{\cal L}= 2i(\bar\kappa_* (1+b
\Gamma^{0}\Gamma_{11})\Gamma^{{0}}e )+{\rm surface\;term}.
\eqno{(3.26)}$$ Using $\Gamma_*$ in (3.6) we see that the  action
(3.23)  is invariant under local unbroken supersymmetries
when $b=-1$. It is also true for $b=1$ if we choose
$\Gamma_*=-\Gamma^{0}\Gamma_{11}$ in (3.6).
\par
The relation of the action of equation (3.25) with ordinary
superparticle action is obtained by considering the equations of
motion for $\varphi^{a'}$ [15] which implies
$$
\varphi^{a'}=-{{{\pi_{0}{}^{a'}}\over{\pi_{0}{}^{0}}}}, \qquad
\pi_{0} {}^ { \underline a} \equiv \dot x^{ \underline
a}+i\bar\theta\Gamma^{ \underline a}\dot \theta. \eqno(3.27)$$
Substituting this algebraic equation back into the action we find
the standard action for the point particle, namely
$$
A= -\int d\tau (\sqrt{-\pi_{0}{}^{ \underline a}\pi_{0}{}^{ \underline b}
\eta_{ \underline a\underline b}}
-i\bar\theta\Gamma_{11}\dot\theta)\;. \eqno(3.28)$$
\par
We will also see now that the kappa symmetry discussed above
coincides with kappa matrix  known in the literature once we
eliminate the non-dynamical Goldstone fields $\varphi_{0}{}^{b'}$.
We rewrite $\kappa_*$ in terms of the 32 component spinor $\kappa$
$$
\bar\kappa_*= \bar\kappa{1\over2}(1+\Gamma^{0}\Gamma_{11}). \eqno{(3.29)}$$
Note that when we work with $\kappa$ instead of $\kappa_*$ we
introduce the reducibility of kappa transformation. This situation
is similar to the non-relativistic branes case [22]
where the kappa symmetry 
is written in terms of the analogous of $\kappa_*$ and it is
therefore irreducible. Let us introduce the field dependent gamma
matrix
$$
\Gamma_\kappa(\varphi)= \tilde\Phi(\varphi)\Gamma_{0}\Gamma_{11}\tilde\Phi
(\varphi)^{-1}= {\Phi_{0}}^{\underline a}\Gamma_{\underline
a}\Gamma_{11}. \eqno{(3.30)}$$ It is trivial to verify 
$\Gamma_\kappa{}^{2}=1$. The transformation of equation (3.22) is then
given by
$$
\delta\bar\theta=\bar\kappa{1\over2}(1+\Gamma^{0}\Gamma_{11})
\tilde\Phi^{-1}=\bar{\tilde\kappa}{1\over2}(1-\Gamma_{\kappa}(\varphi)), \qquad
\tilde\kappa\equiv\tilde\Phi\kappa. \eqno{(3.31)}$$ Once we eliminate
$\varphi^{b'}$ we find that
$$
\Gamma_\kappa(x,\theta)=  {{\pi_{0}{}^{\underline
a}\Gamma_{\underline a} \Gamma_{11}}
\over{\sqrt{-{\pi_{0}{}^{\underline b}}{\pi_{0}{}^{\underline
c}}\eta_ {\underline b \underline c}}}} \eqno{(3.32)}$$ which is the
well known expression of the $\Gamma_\kappa$ matrix [23].
\medskip

Summing up, the action, with the lowest number of derivatives,  for
the super point particle is uniquely determined by the non-linear
realisations once we take the local subalgebra to include the
unbroken supersymmetries and translations. Furthermore we have seen
that the local transformations for the unbroken supersymmetries
reduces to the usual expression for  kappa transformations once we
have eliminated the non-dynamical Goldstone fields corresponding to
the broken Lorentz transformations. However, the latter fields   do
play a crucial role in the construction.

 Here we do not  discuss
the world line diffeomorphism explicitly.
One can prove along the lines of previous section that the
diffeomorphism is equivalent to the $P_0$ transformation
combined with a $Q_*$ transformation.

\medskip
{\bf Super p-brane}
\medskip

We have seen in the bosonic case that the local translations alone are
not an invariance of the p-brane action and  we need to modify them by
adding a local Lorentz transformation whose  parameters are not
independent of the parameters of the local translations. As we will now
see   an analogous situation occurs
for the unbroken supersymmetry transformations of the supersymmetric branes .
The transformations of the Cartan forms under the
local transformations corresponding to $P_a$, $Q^*$ and $J^a{}_{b'}$
are given in (3.16) and (3.17).

An  action of super p-branes which was constructed from the   theory
of non-linear realisation but not taking the unbroken translations
and supersymmetries in the local subalgebra was given by [15]
$$
A=-T\int ({\cal L}^{NG}+b\;{\cal L}^{WZ}) \eqno{(3.33)}$$ where
$$
{\cal
L}^{NG}=-{1\over{(p+1)!}}\epsilon_{a_{0},...,a_p}e^{a_{0}}\wedge
.... \wedge e^{a_p}, \eqno{(3.34)}$$ and the WZ action [24] is given
by
$$
{\cal L}^{WZ}= {-1\over{(p+1)!}} \sum_{r={0}}^p(-1)^r
\left(\matrix{p+1\cr r+1}\right)\pi^{\underline a_p}\wedge \ldots
\wedge \pi^{\underline a_{r+1}}\wedge K^{\underline a_r}\wedge
\ldots \wedge K^{\underline a_1}\wedge K_{\underline a_1,...
\underline a_p}, \eqno{(3.35)}$$ where
$\left(\matrix{p+1\cr r+1}\right)$ 
is the binomial coefficient and
$$ K^{\underline
a}=i\bar\theta\Gamma^{\underline a}d\theta,\qquad K_{\underline
a_1,...\underline a_p}=i\bar\theta \Gamma_{\underline
a_1,...\underline a_p}d\theta \;. \eqno{(3.36)}$$
The action depends on all the Goldstone bosons of the theory
including those corresponding to the broken Lorentz transformation,
namely
$\phi_{a}{}^{b'}$. However, the WZ term ${\cal L}^{WZ}$ does not
depend on these latter fields .
\par
Like in the case of the superparticle let us assume that our brane
breaks half of the supersymmetries and there remain the  unbroken
supersymmetry generators  $Q^*_{\underline \alpha}$.
 If we consider the
unbroken generators as elements of the local algebra $H$, it can be
seen that there is no action of the type equation (3.33) that is
invariant under the right transformations generate by $Q^*$. As in
the diffeomorphism of the bosonic p-brane we should also introduce
local  transformation associated to the broken Lorentz
transformations. Let us  consider the right action on the coset
given by
$$h=e^{\bar\kappa_*Q^*+ir_a{}^{b'}J^a{}_{b'}}\eqno{(3.37)}$$
where the
$\kappa_*$ are 16 independent spinor gauge parameters that depend on
$\xi$ and $r_a{}^{b'}$ are dependent gauge parameters linearly
related to $\kappa_*$. They are the supersymmetric analogous of the
ones appearing in the right action of the unbroken translation
(2.30).

The gauge transformations on the fields are given by given
$$
\delta x^{\underline a}=
i\delta\bar\theta\Gamma^{\underline a}\theta,
\qquad \delta\bar\theta^{\underline
\alpha}=\bar\kappa_*^{\underline \beta} \tilde\Phi^{-1}_{\underline
\beta}{}^ {\underline \alpha}, \qquad \delta
{\phi^{aa'}}=r^{bb'}\left({W\over \sinh W}\right)_{bb'} {}^{aa'}
\eqno{(3.38)}$$
where ${{r_{b'}}^{a}=(\Phi^{-1}\delta\Phi)_{b'}}^{a}$ is 
analogous of the components of the spin connection ${{w_{b'}}^{a}}$
in which the differential $d$ is replaced by the variation $\delta$.
$W$ is defined in the appendix of [15] by
$$
(W^{2}{)_{aa'}}^{bb'}=-(\phi_a{}^{c'}\phi^b{}_{c'})
{\delta_{a'}}^{b'}-{\delta_a}^b {(\phi^c{}_{a'}\phi_c{}^{b'})}
+2{\phi_a}^{b'}{\phi^b}_{a'}. \eqno(3.39)$$

The relation of $r_a{}^{b'}$ and $\kappa_*$ will be determined by
requiring the invariance of the lagrangian under (3.37).
We can compute the variation of the lagrangian under this
transformation
$$
\delta_\kappa{\cal L}^{NG} = {{-1}\over{p!}}\epsilon_{a_{0}...a_p}
e^{a_{0}}...e^{a_{p-1}}(2i(\bar\kappa_*\Gamma^{a_p}e)+e^{b'}
{r_{b'}}^{a_p})
,\qquad
\eqno{(3.40)}
$$
$$
\delta_\kappa{\cal L}^{WZ}= {{(-1)^p}\over{p!}} e^{\underline
a_{0}}...e^{\underline a_{p-1}}( 2i\,\bar\kappa_*\Gamma_{\underline
a_{0}...\underline a_{p-1}}e) +{\rm surface\;term}.
\eqno{(3.41)}$$
$\delta_\kappa{\cal L}^{WZ}$ is separated into two terms, the first
one includes sum of terms with only longitudinal indices $a_j$'s.
Using $\Gamma_*$ in (3.7)  and
$$
\Gamma_{a_{0}...a_{p-1}}=-
\epsilon_{a_{0}...a_{p-1}a_{p}}\Gamma_*\Gamma^{a_p},\qquad
\eqno{(3.42)}$$ {the} sum of the first terms of the
$\delta_\kappa{\cal L}^{WZ}$ and the $\delta_\kappa{\cal L}^{NG}$
becomes
$$
\delta_\kappa{\cal L}_1: =
{{-2i}\over{p!}}\epsilon_{a_{0}...a_p}
e^{a_{0}}...e^{a_{p-1}}\left(\bar\kappa_*(1+b{(-1)^p}\Gamma_*)\Gamma^{a_p}e
\right).
\eqno{(3.43)}$$
Since $\bar\kappa_*=\bar\kappa_*{1\over 2}(1+\Gamma_*)$ (3.43) vanishes when
$b=(-1)^{p+1}$.
(If we were taken the opposite sign choice for $\Gamma_*$, the lagrangian with
$b={(-1)^p}$ is invariant.)
The remaining terms in the variation of the lagrangian are
$$
\delta_\kappa{\cal L}_2: =
{{-1}\over{p!}}\epsilon_{a_{0}...a_p}
e^{a_{0}}...e^{a_{p-1}}e^{b'}
{r_{b'}}^{a_p}+b{{(-1)^p}\over{p!}}\sum{}' e^{\underline
a_{0}}...e^{\underline a_{p-1}} (2i\bar\kappa_*\Gamma_{\underline
a_{0}...\underline a_{p-1}}e). \eqno{(3.44)}$$ Here the sum
$\sum{}'$ does not include terms with only longitudinal indices
$a_j$'s thus at least one of ${\underline a_j}$ is transverse
$a'_j$. Since the  world volume vielbein ${e_i}^a$ is non-singular
we can determine ${r_{b'}}^{a_p}$ in terms of $\kappa_*$ from
$\delta_\kappa{\cal L}_2={0}$. Thus the kappa transformation of the
Goldstone fields ${\varphi_{b}{}^{a'}}$ is determined from (3.44).
The total lagrangian is pseudo-invariant under the kappa
transformation.

Let us see now that the kappa symmetry discussed above coincides
with kappa transformation known in the literature once we eliminate
the non-dynamical Goldstone fields $\varphi_{a}{}^{b'}$.
 We write  $\bar\kappa_*=\bar\kappa{\cal P}$ in
term of  $\bar\kappa$ spinor with independent components
$$
\bar\kappa_*=\bar\kappa{1\over2}(1+\Gamma_*).
\eqno{(3.45)}$$ In terms of
this kappa parameter the transformation {(3.38)} is given by
$$
\delta\bar\theta=\bar\kappa{1\over2}(1+\Gamma_*)\tilde\Phi^{-1} =
\bar{\tilde\kappa}{1\over2}(1+\Gamma_{\kappa}), \qquad
\tilde\kappa\equiv\tilde\Phi\kappa, \eqno{(3.46)}$$ where
$$
\Gamma_\kappa(\varphi)= \tilde\Phi\Gamma_*\tilde\Phi^{-1}=
{{1}\over{(p+1)!}}\epsilon^{a_{0}...a_p}{\Phi^{-1}_{a_{0}}}^{\underline
b_{0}}... {\Phi^{-1}_{a_p}}^{\underline b_p}\Gamma_{\underline
b_{0}}...\Gamma_ {\underline b_p} \eqno{(3.47)}$$ that verifies
$\Gamma_\kappa{}^{2}=1$. Once we eliminate $\varphi_{a}{}^{b'}$ we get
$$
\Gamma_\kappa(x,\theta)= {{1}\over{(p+1) !\sqrt{-\det
G_{ij}}}}\epsilon^{i_{0}...i_p} {\pi_{i_{0}}}^{\underline b_{0}}...
{\pi_{i_p}}^{\underline b_p}\Gamma_{\underline b_{0}}...\Gamma_
{\underline b_p}. \eqno{(3.48)}$$ which is the well known expression
of the $\Gamma_\kappa$ matrix [25-26].

\medskip
Summing up, for the case of the super p-brane we have seen that it
is not enough to consider the unbroken supersymmetries in the local
subgroup of the  non-linear realisation in order to have the  invariance of
the super-brane action. We need to also carry out a local   transformation
associated with the Lorentz broken generator with the parameters
${r_{b'}}^{a}$ as functions of
$\kappa_*$.

\medskip
{\bf 4. Discussion. }
\medskip
In this paper we have derived the dynamics of bosonic and
superbranes using a theory of non-linear realisations in which  the
unbroken translations and supercharges are part of the respective
local sub-algebras. For the bosonic point particle and the
superpoint particle this has lead to a group theoretic derivation of
world line reparameterisation invariance and kappa symmetry
respectively. However, for the bosonic p-brane and super p-brane we
have  had to supplement the local transformations by a field
dependent broken Lorentz  transformation. The latter step is outside of
the non-linear realisation as formulated in this paper. It might
seem as if it were a compensating transformation, however, as
formulated in this paper the unbroken Lorentz transformations have
no local symmetry associated with them. It may be that there is a
more general formulation of the non-linear realisation in which this
field dependent broken Lorentz transformation is part of the
non-linear realisation.
\par
One of the aims of this paper is to further develop the theory of
non-linear realisations in the context of branes so that one might
apply it to the  conjectures in reference [27] concerning brane
dynamics and the Kac-Moody algebra $E_{11}$.
\medskip

{\bf Acknowledgements}

We acknowledge discussions with Andr\'es Anabal\'on, Jaume Gomis,
Norisuke Sakai, Toine Van Proeyen, Paul Townsend, Jorge Zanelli.
Joaquim Gomis and Peter West would like to thank the CECS, Valdivia,
Chile, where part of this work was carried out,  for their support
and hospitality.   We also thank Benasque Center of Science for
their hospitality. This work is supported in part by the European
EC-RTN network MRTN-CT-2004-005104, MCYT FPA 2004-04582-C02-01,
CIRIT GC 2005SGR-00564. PW is supported by a PPARC senior fellowship
PPA/Y/S/2002/001/44. This work was in addition supported in part by
the PPARC grant PPA/G/O/2000/00451 and the EU Marie Curie research
training work grant HPRN-CT-2000-00122.

\medskip
 {\bf References}
\medskip
\noindent {  }
\item{[1]}
   S.~Weinberg,
   {\it ``Nonlinear Realizations Of Chiral Symmetry,''}
   Phys.\ Rev.\  {\bf 166} (1968) 1568; 
  J.~S.~Schwinger,
  {\it``Chiral Dynamics,''}
  Phys.\ Lett.\ B {\bf 24}(1967) 473;
  {\it``Partial Symmetry,''}
  Phys.\ Rev. \ Lett.  {\bf 18}(1967) 923; 
  D.~B.~Fairlie and K.~Yoshida,
  ``Chiral symmetry from divergence requirements,''
  Phys.\ Rev.\  {\bf 174} (1968) 1816.

\item {[2]}
     S.~R.~Coleman, J.~Wess and B.~Zumino,
     {\it ``Structure Of Phenomenological Lagrangians. 1,''}
     Phys.\ Rev.\  {\bf 177} (1969) 2239,
C.~G.~Callan, S.~R.~Coleman, J.~Wess and B.~Zumino,
     {\it ``Structure Of Phenomenological Lagrangians. 2,''}
     Phys.\ Rev.\  {\bf 177} (1969) 2247.
\item{[3]}
  V.~P.~Akulov and D.~V.~Volkov,
  {\it ``Goldstone fields with spin 1/2''}
  Teor.\ Mat.\ Fiz.\  {\bf 18} (1974) 39;
  {\it ``Gauge fields for symmetry group with spinor parameters,''}
  Teor.\ Mat.\ Fiz.\  {\bf 20} (1974) 291.
\item{[4]}
  A.~Salam and J.~A.~Strathdee,
  {\it ``On Superfields And Fermi-Bose Symmetry,''}
  Phys.\ Rev.\ D {\bf 11} (1975) 1521.
\item{[5]}
  A.~Salam and J.~A.~Strathdee,
  {\it ``Nonlinear realizations. 2. Conformal symmetry,''}
  Phys.\ Rev.\  {\bf 184} (1969) 1760;
  C.~J.~Isham, A.~Salam and J.~A.~Strathdee,
   {\it ``Nonlinear realizations of space-time symmetries. Scalar and tensor
  gravity,''}
  Annals Phys.\  {\bf 62} (1971) 98.

\item{[6]}
     A.~P.~Balachandran, A.~Stern and C.~G.~Trahern,
     {\it ``Nonlinear Models As Gauge Theories,''}
     Phys.\ Rev.\ D {\bf 19} (1979) 2416.

\item{[7]}
     M.~Bando, T.~Kugo, S.~Uehara, K.~Yamawaki and T.~Yanagida,
     {\it ``Is Rho Meson A Dynamical Gauge Boson Of Hidden Local
Symmetry?,''}
     Phys.\ Rev.\ Lett.\  {\bf 54} (1985) 1215.

\item{[8]} J. Hughes and J.  Polchinski, {\it Partially broken global
supersymmetry and the superstring}, Nucl Phys  B278 (1986) 147.

\item{[9]} J. Gauntlett, K. Itoh, and P. Townsend, {\it Superparticle
with extrinsic curvature}, Phys. Lett. B238, (1990) 65.

\item{[10]}    J.~P.~Gauntlett, J.~Gomis and P.~K.~Townsend,
      {\it ``Particle Actions As Wess-Zumino Terms For Space-Time
(Super) Symmetry
      Groups,''}
      Phys.\ Lett.\ B {\bf 249} (1990) 255;
  J.~P.~Gauntlett and C.~F.~Yastremiz,
 {\it``Massive Superparticle in D = (2+1) and Sigma Model Solitons,''}
  Class.\ Quant.\ Grav.\  {\bf 7} (1990) 2089.

\item{[11]}  J. Bagger, A. Galperin,
{\it
     ``Matter couplings in partially broken extended supersymmetry,''}
Phys. Lett. {\bf B 336} (1994) 25; {\it
     ``A new Goldstone multiplet for partially broken supersymmetry,''}
Phys. Rev. {\bf D 55} (1997) 1091; {\it ``The tensor Goldstone
multiplet for partially broken supersymmetry,''} Phys. Lett. {\bf B
412} (1997) 296.

\item {[12]} S. Bellucci, E. Ivanov, S. Krivonos, {\it ``Partial
breaking of N = 1 D = 10 supersymmetry,''} Phys. Lett. {\bf B 460}
(1999) 348; E. Ivanov, S. Krivonos, {\it ``N = 1 D = 4 supermembrane
in the coset approach,''} Phys. Lett. {\bf B 453} (1999) 237; S. V.
Ketov, {\it ``Born-Infeld-Goldstone superfield actions for
gauge-fixed D-5 and D-3 branes in 6d,''} { Mod. Phys. Lett.\/} {\bf
A14} (1999) 501.

\item{[13]}   P. West,   {\it ``Automorphisms, non-linear realizations
and branes,''}
      JHEP {\bf 0002} (2000) 024

\item{[14]} E.~A.~Ivanov and V.~I.~Ogievetsky,
      {\it ``The Inverse Higgs Phenomenon In Nonlinear Realizations,''}
      Teor.\ Mat.\ Fiz.\  {\bf 25} (1975) 164.

\item{[15]}
  J.~Gomis, K.~Kamimura and P.~West,
   {\it ``The Construction Of Brane And Superbrane Actions Using Non-Linear
  Realisations,''}
  arXiv:hep-th/0607057.

\item{[16]}
      J.~Brugues, T.~Curtright, J.~Gomis and L.~Mezincescu,
      {\it ``Non-relativistic strings and branes as non-linear
realizations of  Galilei
      groups,''}
      Phys.\ Lett.\ B {\bf 594} (2004) 227.

\item{[17]}
      J.~Brugues, J.~Gomis and K.~Kamimura,
      {\it ``Newton-Hooke algebras, non-relativistic branes and
generalized pp-wave
      metrics,''}
      Phys.\ Rev.\ D {\bf 73} (2006) 085011.

  \item {[18]}
  I.~A.~Bandos and A.~A.~Zheltukhin,
  {\it ``Green-Schwarz Superstrings In Spinor Moving Frame
  Formalism,''}
  Phys.\ Lett.\ B {\bf 288} (1992) 77,
  I.~A.~Bandos, D.~P.~Sorokin, M.~Tonin, P.~Pasti and D.~V.~Volkov,
   {\it ``Superstrings and supermembranes in the doubly supersymmetric geometrical
  approach,''}
  Nucl.\ Phys.\ B {\bf 446} (1995) 79.

\item {[19]}I.~N.~McArthur,
  {\it ``Kappa-symmetry of Green-Schwarz actions in coset
  superspaces,''}
  Nucl.\ Phys.\ B {\bf 573} (2000) 811.

\item{[20]}
      A.~Achucarro, J.~M.~Evans, P.~K.~Townsend and D.~L.~Wiltshire,
      {\it ``Super P-Branes,''}  Phys.\ Lett.\ B {\bf 198} (1987) 441.

\item{[21]}
     J.~A.~De Azcarraga and P.~K.~Townsend,
     {\it ``Superspace Geometry And Classification Of Supersymmetric
Extended
      Objects,''}
      Phys.\ Rev.\ Lett.\  {\bf 62} (1989) 2579.

\item{[22]}
  J.~Gomis, K.~Kamimura and P.~K.~Townsend,
  {\it``Non-relativistic superbranes,''}
  JHEP {\bf 0411} (2004) 051,
  J.~Gomis, J.~Gomis and K.~Kamimura,
  {\it ``Non-relativistic superstrings: A new soluble sector of 
$AdS_5\times S^5$,''}
  JHEP {\bf 0512} (2005) 024

\item{[23]}
  E.~Bergshoeff and P.~K.~Townsend,
  {\it ``Super D-branes,''}
  Nucl.\ Phys.\ B {\bf 490} (1997) 145.

\item{[24]}
     J.~M.~Evans,
     {\it ``Super P - Brane Wess-Zumino Terms,''}
     Class.\ Quant.\ Grav.\  {\bf 5} (1988) L87.

\item{[25]}
  J.~Hughes, J.~Liu and J.~Polchinski,
  {\it ``Supermembranes,''}
  Phys.\ Lett.\ B {\bf 180} (1986) 370.

\item{[26]}
  E.~Bergshoeff, E.~Sezgin and P.~K.~Townsend,
  {\it ``Supermembranes And Eleven Dimensional Supergravity,''}
  Phys.\ Lett.\ B {\bf 189} (1987) 75.

\item{[27]}  P. West, {\it "$E_{11}$ origin of Brane charges
and U- duality multiplets,"}
 JHEP {\bf 0408} (2004) 052.

\end